\documentclass[a4paper,11pt]{article}
\usepackage{pos}
\usepackage{wrapfig}
\usepackage[percent]{overpic}

\title{Deep-learning event reconstruction for the Cherenkov Telescope Array Observatory with CTLearn}
\ShortTitle{Event reconstruction with CTLearn for the CTAO}

\author*[a]{ 
    B. Lacave
}
\author[a]{ 
    T. Miener
}
\author[b]{ 
    A. Cerviño
}
\onbehalf{on behalf of the CTAO-LST Project}

\affiliation[a]{Département de Physique Nucléaire et Corpusculaire, Université de Genève, Faculté de Sciences,\\
1211 Genève 4, Switzerland}
\affiliation[b]{IPARCOS-UCM, Instituto de Física de Partículas y del Cosmos, and EMFTEL Department, Universidad Complutense de Madrid,\\ Plaza de Ciencias, 1. Ciudad Universitaria, 28040 Madrid, Spain}

\emailAdd{bastien.lacave@unige.ch}

\abstract{The Cherenkov Telescope Array Observatory (CTAO) is the next-generation ground-based observatory for very-high-energy (VHE) gamma-ray astronomy. The Large-Sized Telescope prototype, LST-1, located on the Canary Island of La Palma, is responsible for observation of the low-energy range of the VHE gamma-ray spectrum. It is undergoing commissioning and has already observed the Crab Nebula as a standard reference source. Accurate reconstruction of shower parameters (e.g. energy, direction, and particle type) is crucial for achieving the scientific goals of the CTAO.
In this work, we use CTLearn to implement deep-learning event reconstruction, as an alternative to the standard Random Forest method. CTLearn is built to be fully compatible with ctapipe, a framework for prototyping the low-level data processing algorithms for the CTAO, and can be seamlessly used for data analysis without changing the general framework. It implements convolution-neural-network based models that take the integrated charge and the relative peak time of calibrated pixels in cleaned images as an input, to infer the primary particle’s properties.
Using Crab Nebula observations as a validation sample, we explore two different approaches. The first is to train a model with Monte-Carlo (MC) simulations covering all possible altitude-azimuth coordinates of the Crab Nebula sample observations, resulting in a single model that can be used to reconstruct events from any Crab Nebula observations. The second approach is to train 10 models along this coordinate line, each incorporating a range of \textasciitilde10° in altitude.
In this contribution, we present our investigation of the performance of CTLearn models, and highlight the potential of CTLearn for future data analysis in the CTAO.}

\ConferenceLogo{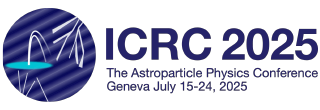}

\FullConference{39th International Cosmic Ray Conference (ICRC2025)\\
15–24 July 2025\\
Geneva, Switzerland\\}

\begin{document}
\maketitle

\section{Introduction}
The Cherenkov Telescope Array Observatory (CTAO)~\cite{2019APh...111...35A} is the next-generation ground-based gamma-ray observatory. It consists of two arrays in La Palma, Spain, and Cerro Paranal, Chile, with numerous Imaging Atmospheric Cherenkov Telescopes (IACTs) of various sizes. A primary objective of the CTAO is to achieve a sensitivity five to ten times greater than current instruments, with energy coverage from 20 GeV to over 300 TeV. This range is accomplished using three IACT types: the Large-, Medium-, and Small-Sized Telescopes (LST, MST, and SST). The LSTs are engineered for the lowest energies (\textasciitilde20 GeV to several hundred GeV), using large 23-meter mirrors and fast electronics to detect the faint Cherenkov light from low-energy particle showers. The CTAO-North site will utilize four LSTs; the first, LST-1, is fully constructed and is now in its commissioning phase, acquiring science-grade data on astrophysical sources.

IACTs like the LST-1 detect Cherenkov photons emitted by the interaction of very high-energy (VHE) particles with the atmosphere (photons, protons, electrons, ...). The resulting pool of light is collected at 1 GHz by a photomultiplier (PMT) camera. The critical task in gamma-ray astronomy is to infer the primary particle's properties (particle type, energy, and direction) from this light. This process, called event reconstruction, is traditionally performed with a machine learning method known as Random Forest (RF). The RF method uses a parameterization of the camera image known as Hillas parameters~\cite{1985ICRC....3..445H}, which includes temporal information, to train decision trees.

This work proposes a deep learning approach to event reconstruction, utilizing deep convolutional neural networks (CNNs) to improve upon established methods. Unlike the RF method, CNNs do not rely on image parametrization. Instead, they use the full information from the integrated charge and relative time development images to learn and extract features directly.

\section{Methodology and data}
CTLearn\footnote{\href{https://github.com/ctlearn-project/ctlearn}{https://github.com/ctlearn-project/ctlearn}}~\cite{miener_2025_15065761} is a deep learning python package developed for performing event reconstruction on IACTs. It uses a range of possible inputs such as waveforms, calibrated or uncalibrated images, etc., to reconstruct events. In the case of this work, the models are fed with 2 images per event (pre-processed into a square-pixel lattice with the \texttt{DL1-Data-Handler}\footnote{\href{https://github.com/cta-observatory/dl1-data-handler}{https://github.com/cta-observatory/dl1-data-handler}}~\cite{tjark_miener_2025_15422957}), the cleaned and calibrated integrated charge image, and the relative peak time image (time at which each pixel reached max amplitude in the waveform). These images are convoluted with various kernels to extract features, that are then passed to a shallow residual neural network (ResNet)~\cite{2015arXiv151203385H} with 33 layers inspired by the Thin-ResNet~\cite{2019arXiv190210107X}, that learns the relevance of each feature~\cite{Goodfellow-et-al-2016} and infers properties of the events.

In order to efficiently conduct the experiments of this work, a companion package called CTLearn Manager\footnote{\href{https://github.com/ctlearn-project/ctlearn}{https://github.com/BastienLacave/CTLearn-Manager}} was developed. It enables the simple training, testing, data prediction, and full benchmarking of the models through a simplified interface. Model collections can be created to compare strategies and experiments, also to other reconstruction algorithms. All plots shown in this work were produced by the CTLearn Manager.

Note on vocabulary: In the following, the term \textbf{model} will define a collection of three models, one for the energy regression task, one for the direction regression task and one for the particle type classification task, unless explicitly stated otherwise. The term tri-model may also be used for clarity. Moreover, the term \textbf{node} will not define parts of the neural networks used in this work, but the pointing position of the telescope in the sky, where simulated showers come from, making for discrete points, for training and testing the models.

\begin{wrapfigure}{r}{0.5\textwidth}
    \centering
    \includegraphics[width=7cm]{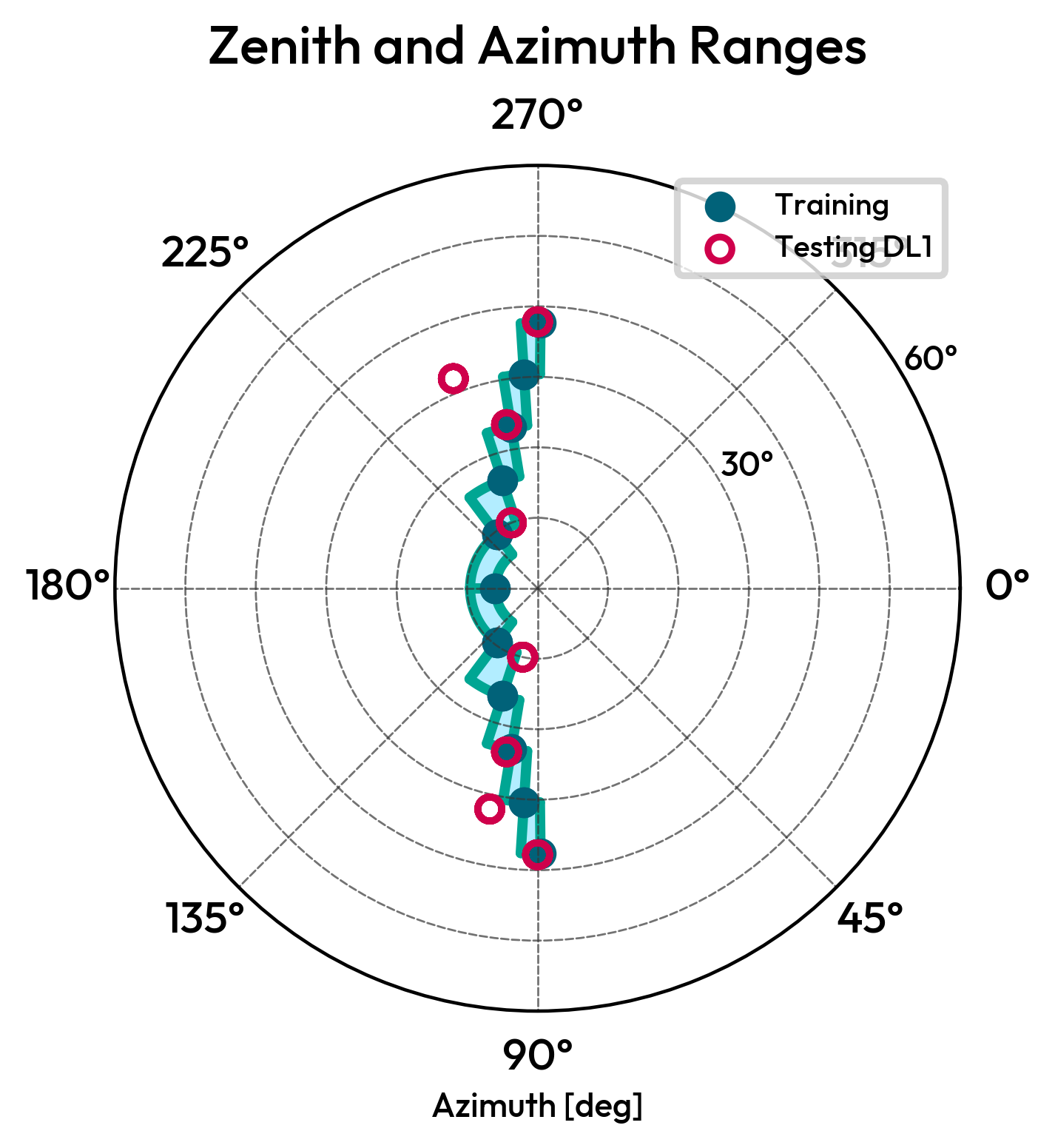}
    \caption{Sky map of the training nodes of the multi models. Each incorporates 2 adjacent nodes. The teal areas represent the ranges of application of each model. Red circles represent the nodes used for testing at different positions along the declination line. For future reference, the models are indexed from 0 to 9 starting at the top (azimuth 270°) and going down towards azimuth 90°.}
    \label{fig:training_sky_map}
\end{wrapfigure}

Two distinct deep-learning strategies were implemented to evaluate the trade-offs between a generalized versus a specialized approach to event reconstruction. The first, labeled the "Single Model", is a universal tri-model trained on Monte-Carlo simulations covering all potential observation directions for the Crab Nebula. This approach offers significant operational simplicity, as a single, robust model can be applied to any dataset without regard to the specific pointing direction. The second strategy, the "Multi-Model" approach, involves training 10 separate tri-models, each specialized for a narrow range of observation altitudes by using only two adjacent training nodes. Although this approach is more computationally intensive and complex to manage, it was hypothesized to yield higher performance by learning direction-specific features. The models are indexed from 0 to 9, covering an azimuth range from 270° to 90° (Figure~\ref{fig:training_sky_map}). Note that the models are not fed with the coordinates of the events.

This project constitutes a total of 11 tri-models, one single model and 10 multi models, each being a group of three CNN models for each of the three reconstruction tasks, resulting in 33 reconstruction tasks to train. The bulk of the training was mad at the \textit{Centro Svizzero di Calcolo Scientifico} (CSCS) in Lugano, Switzerland, on NVIDIA GH200 Grace Hoppers, allowing for under 6h training for each multi model for 7 epochs. Due to a larger amount of training data for the single model, a training time of \textasciitilde2 days was needed. The Monte-Carlo (MC) data was produced using \texttt{CORSIKA}~\cite{Heck:1998vt} and \texttt{sim\_telarray}~\cite{Bernlohr:2008kv}. Each models is trained on diffuse gamma simulations (events simulated in a cone with a given field of view), in addition to diffuse protons for the classification task. Due to the lower amount of protons in each node, the adjacent nodes were also included in the training, as well as the 4 nodes with same zenith but opposite azimuth, in order to have a balanced training set between gamma and proton samples. 

Finally, the real LST-1 Crab nebula data processed in this work is a subset of the LST-1 Crab performance paper sample~\cite{Abe_2023}, including the following runs 2933, 2934, 2974, 2975, 2976, 2989, 2990, 2929, 2914, 2949, 2968, 2969, 3093, 3094, 3271, 3272, 6304, accounting for 5h of effective time.

\section{Monte-Carlo performance}

\begin{figure}[h]
    \centering
    \includegraphics[width=4.9cm]{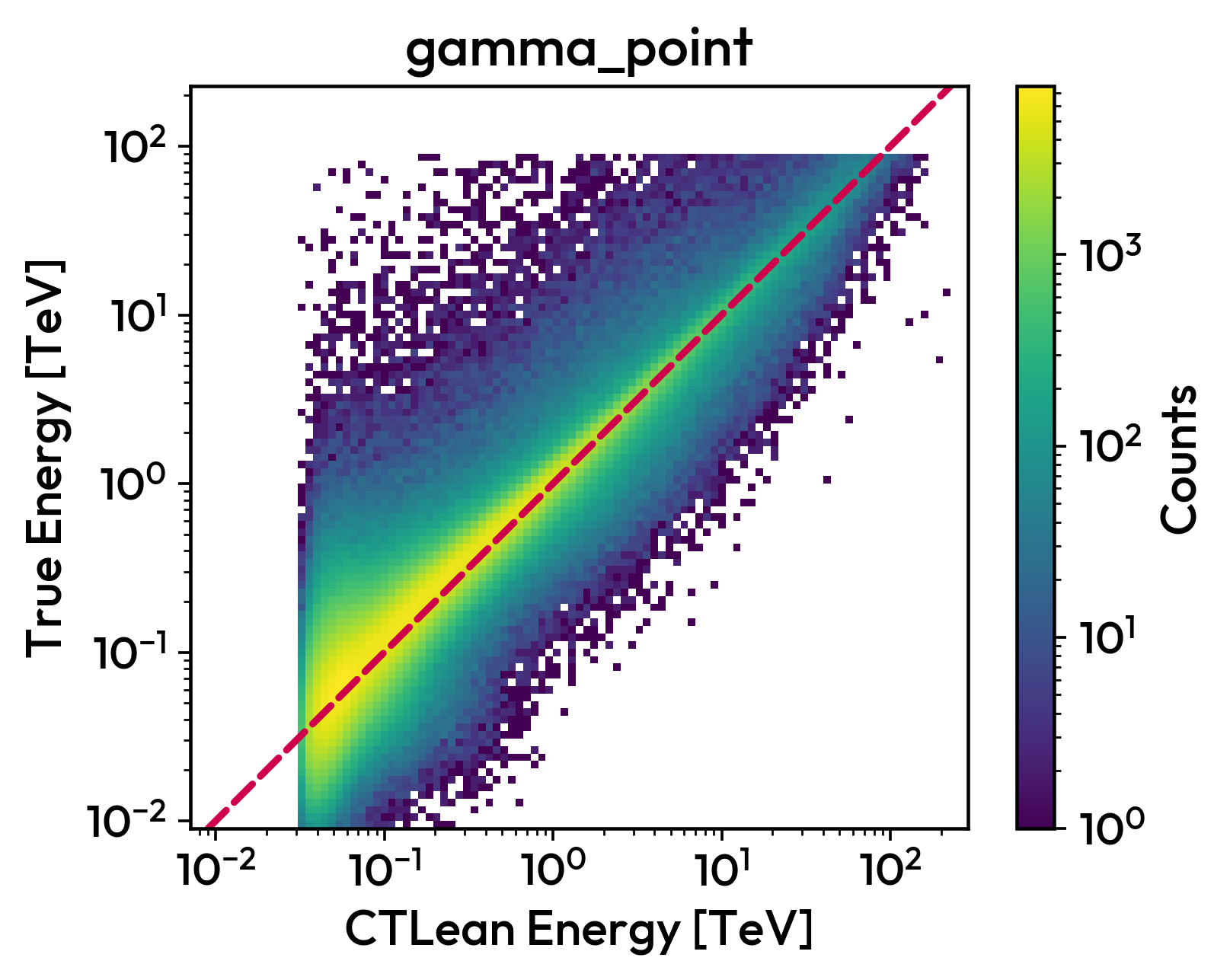}
    \includegraphics[width=4.9cm]{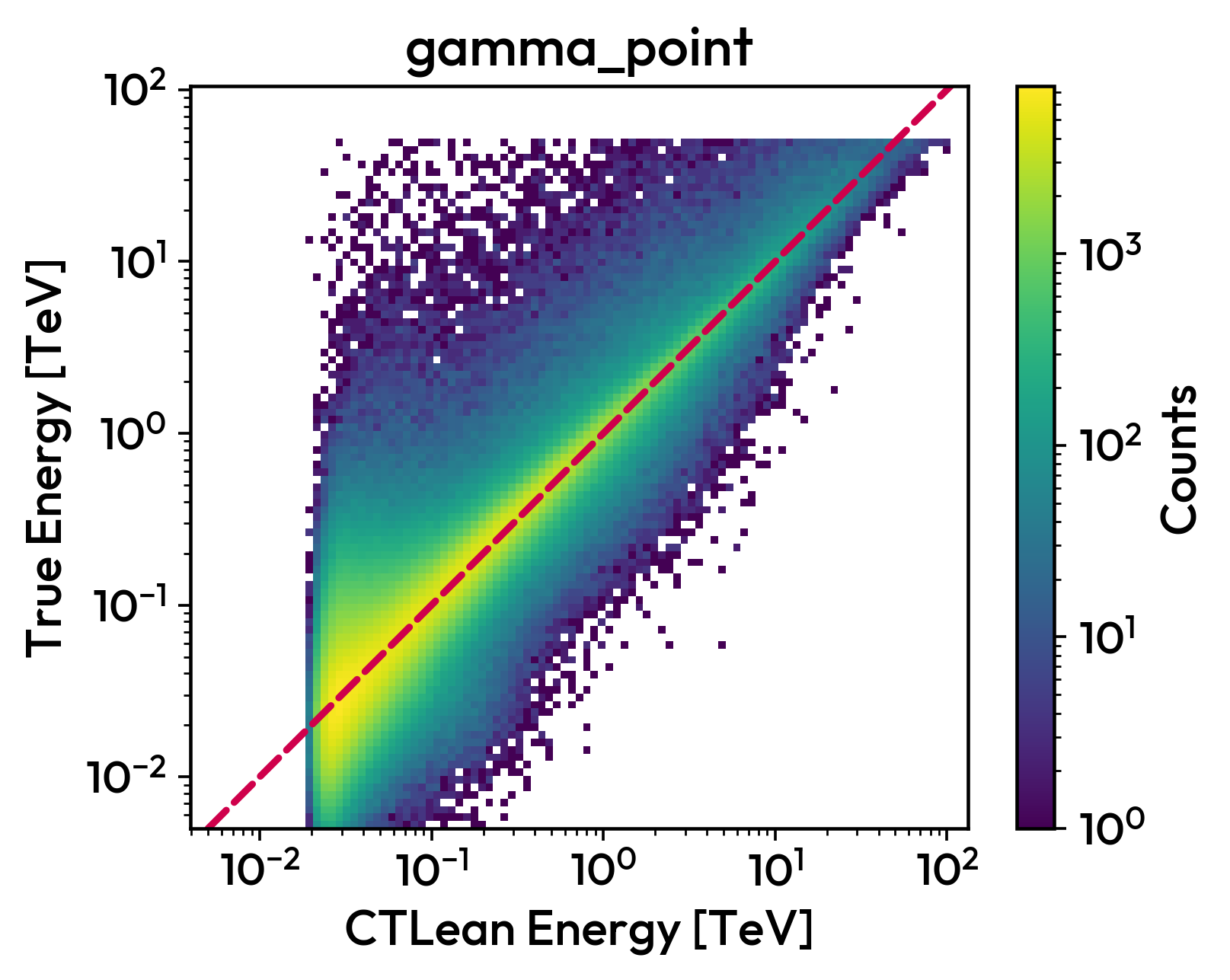}
    \includegraphics[width=4.9cm]{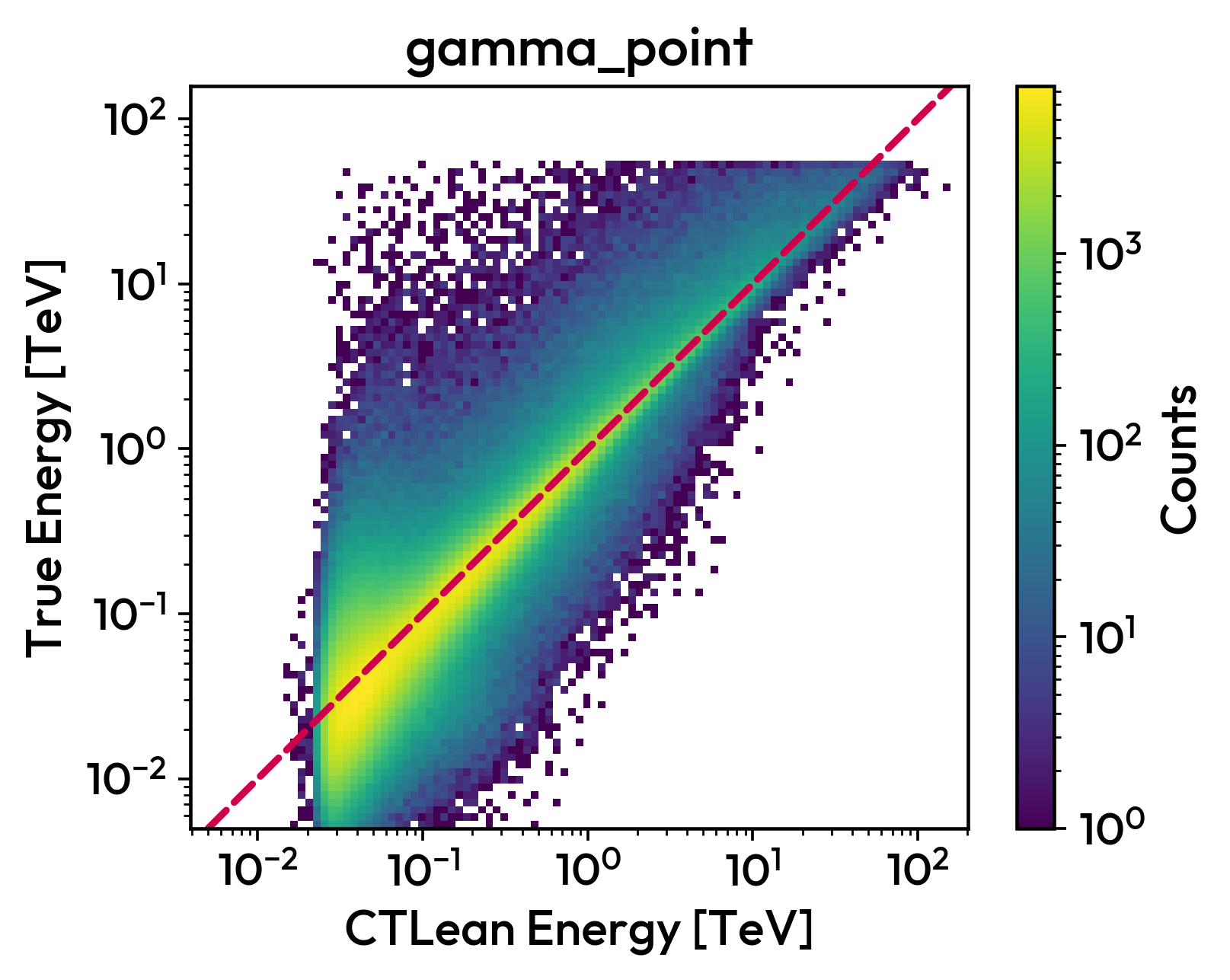}
    \caption{(\textbf{Left}): Migration matrix for the multi model n°0 tested at zenith 37.814° and azimuth 270.0°. (\textbf{Center}): Migration matrix for the multi model n°4 tested at zenith 10.0° and azimuth 248.117°. (\textbf{Right}): Migration matrix for the single model tested at zenith 10.0° and azimuth 248.117°.}
    \label{fig:migration_mtx}
\end{figure}

\begin{figure}[h]
    \centering
    \includegraphics[width=4.9cm]{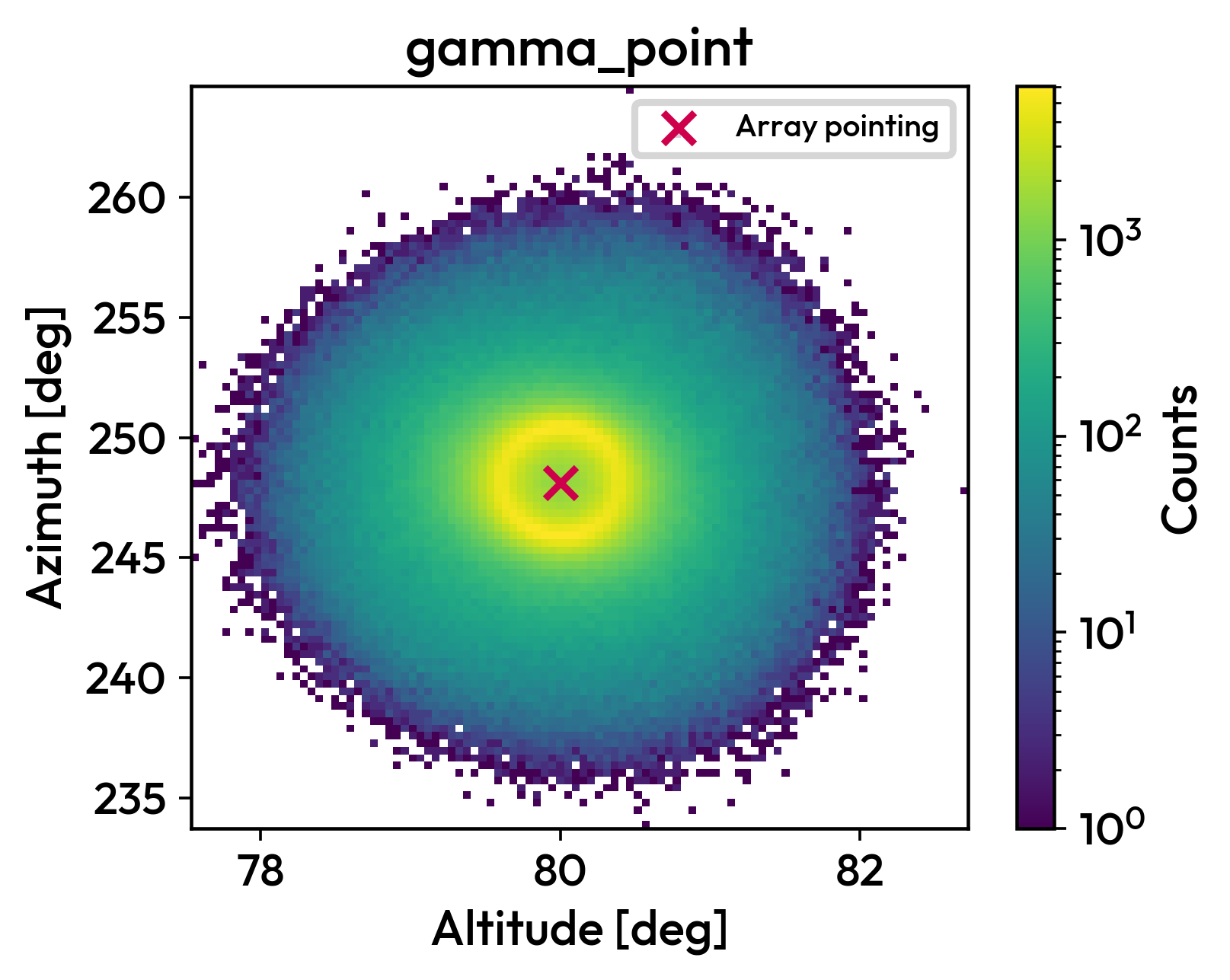}
    \includegraphics[width=4.9cm]{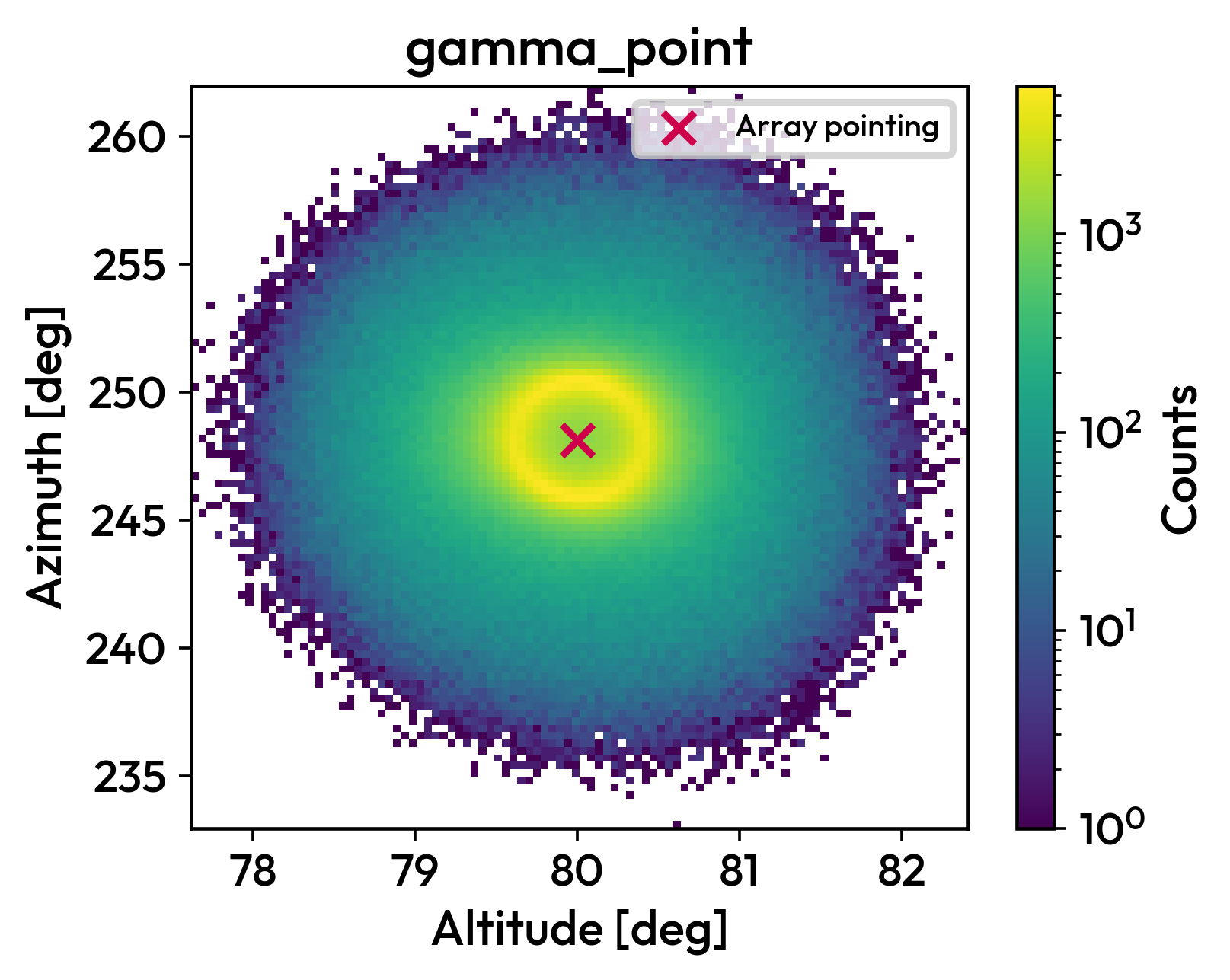}
    \includegraphics[width=4.9cm]{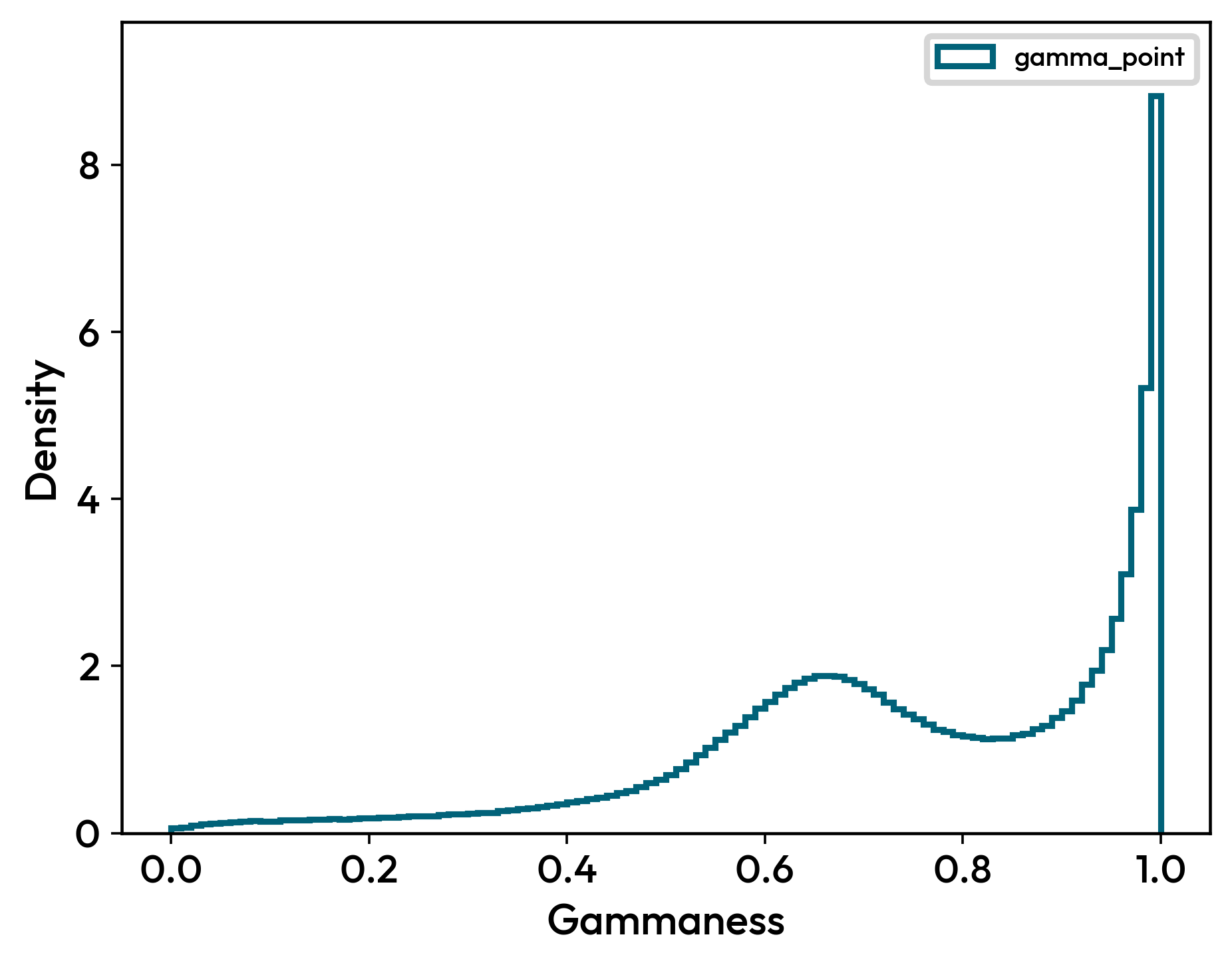}
    \caption{(\textbf{Left}): Sky map of reconstructed event positions for the multi model n°4 tested at zenith 10.0° and azimuth 248.117°. (\textbf{Center}): Sky map of reconstructed event positions for the single model tested at zenith 10.0° and azimuth 248.117°. (\textbf{Right}): Gammaness distribution for Multi Model n°4 for the testing point-like gamma sample.}
    \label{fig:mc_skymaps}
\end{figure}

Trained models are tested on ring-wobble gammas (as seen on Figure \ref{fig:mc_skymaps}) from the closest testing node along the Crab nebula declination line. A series of low-level checks are performed to ensure the models work as expected. The migration matrices for both the multi-model and single-model approaches (Figure \ref{fig:migration_mtx}) show a high concentration of events along the identity line, where reconstructed energy equals true Monte-Carlo energy. This confirms that both models accurately reconstruct event energies across the tested range. Similarly, the sky maps of reconstructed event positions (Figure \ref{fig:mc_skymaps}) demonstrate that reconstructed event directions are tightly clustered around the true source position for both model types, indicating precise angular reconstruction. Finally, due to the lack of testing protons data, the classifier could only be tested on gammas. Their distribution of gammaness is shown on Figure \ref{fig:mc_skymaps}.

Performance on MC data is evaluated on an optimized sample of the data, generally involving a cut on the gamma-hadron classifier value, also called gammaness. By keeping event with high gammaness (probability of being a gamma event), one obtains the performance for gamma-like events. Moreover a cut of the angular separation from the source (theta) is applied to benchmark energy resolution on the target events, discarding outliers coming from background regions. This cut is referred to as the theta cut.
In order to achieve the best performance possible on real data, cuts are optimized in an energy-dependent way on MC, such that in each energy bin, a certain percentage of the gamma events are retained. We call this percentage efficiency, and it is defined for both gammaness and angular separation (theta efficiency). This way, cuts are obtained such that a given percentage of gamma events (on MC) are kept, en then, a given percentage of those with a theta cut. Note that for the angular resolution benchmarking, no theta cuts are applied as this would defeat the purpose of evaluating the capacity of the model to infer the position accurately, and artificially improve it. These cuts can then be applied of real data, and given that the MC matches the data, the same percentages of gamma-like events will be kept in the analysis, eliminating proton-like events.
For cuts optimization, the \texttt{ctapipe-optimize-event-selectio} tool is used, then IRFs (Instrument Response Functions) are computed using these cuts with \texttt{ctapipe-compute-irf}~\cite{karl_kosack_2021_4581045}.

\begin{figure}[b]
    \centering
    \begin{overpic}[width=7.45cm]{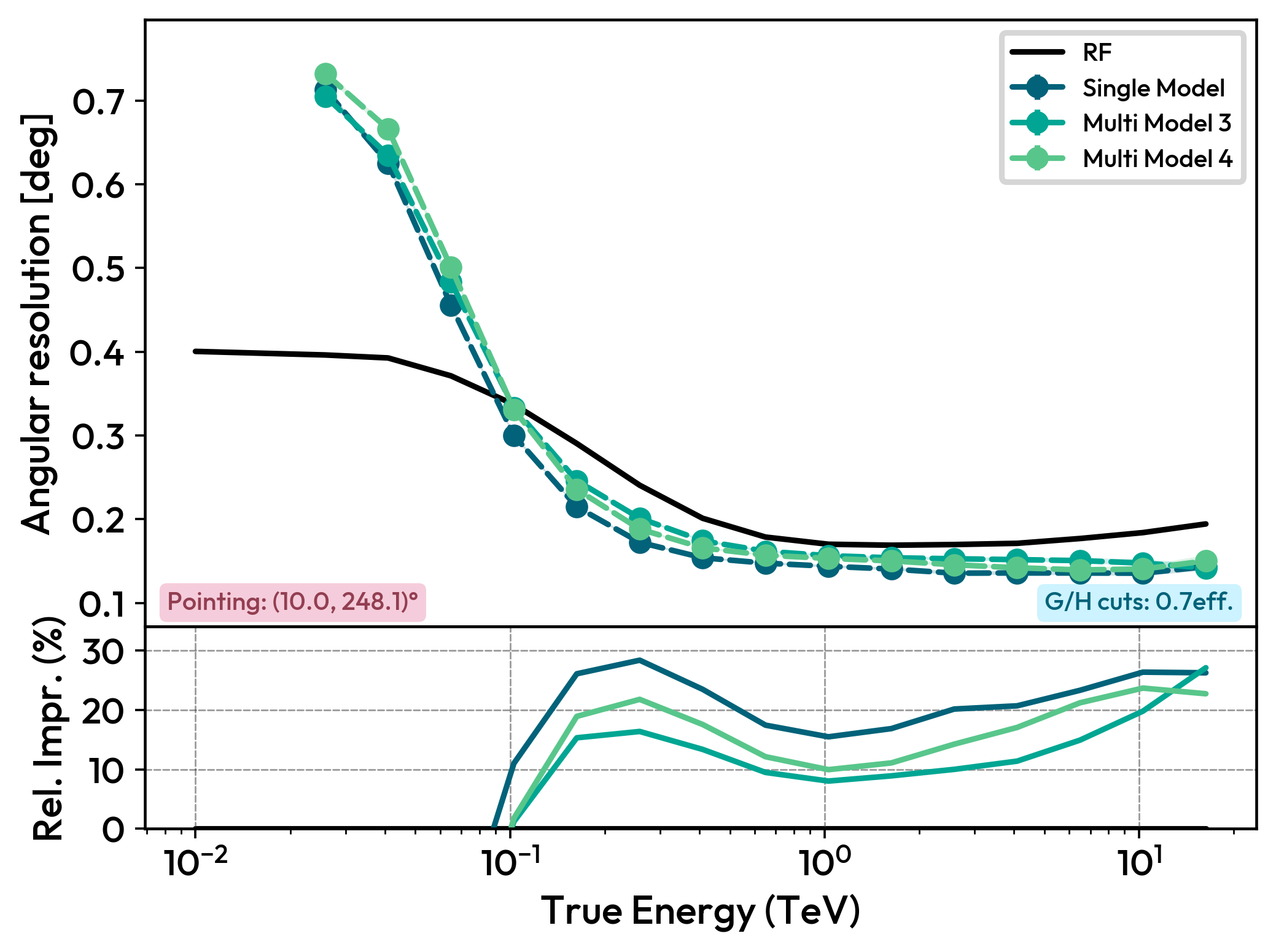}
        \put(45,60){\textcolor{gray}{Preliminary}}
    \end{overpic}
    \begin{overpic}[width=7.45cm]{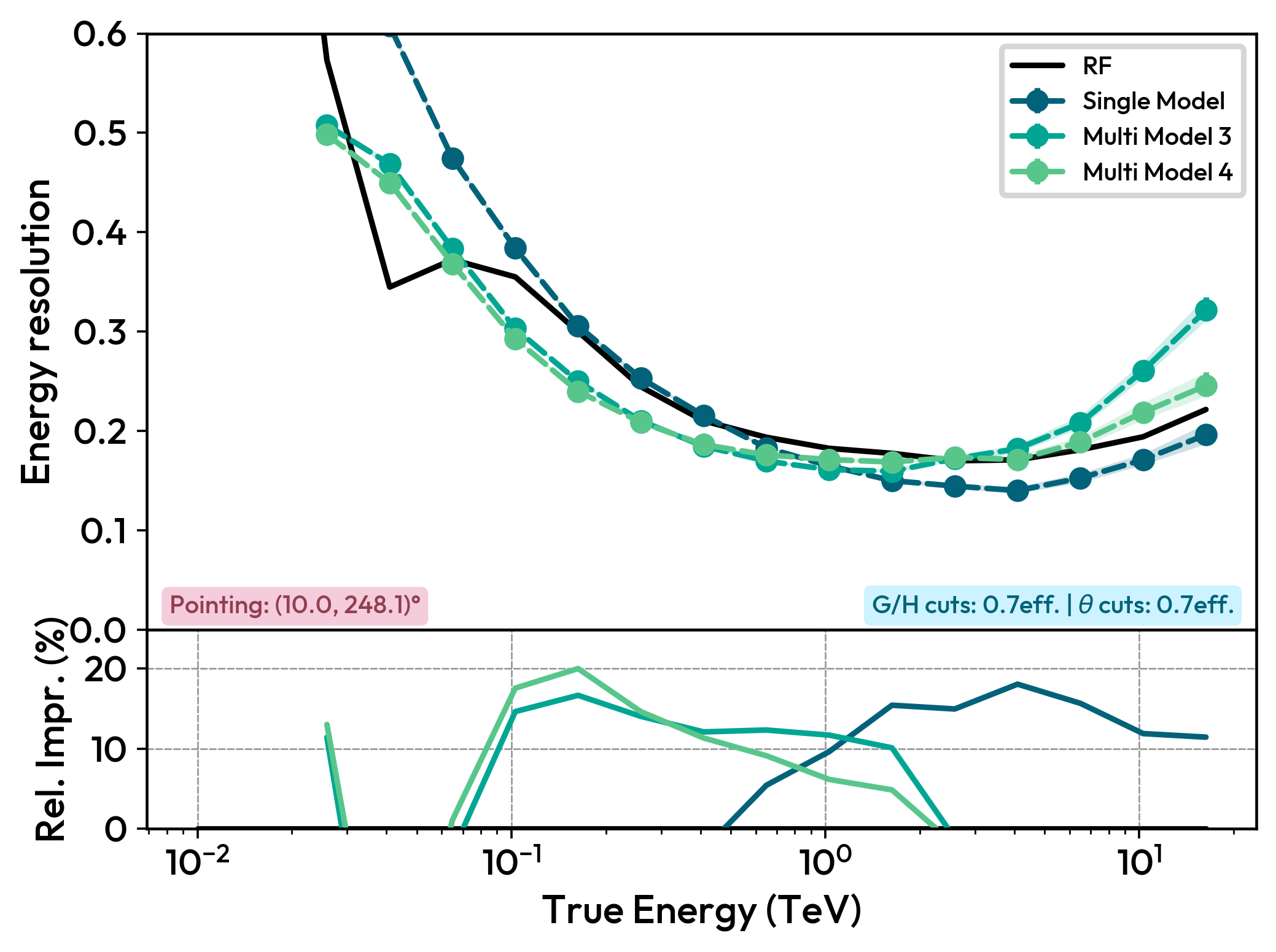}
        \put(45,60){\textcolor{gray}{Preliminary}}
    \end{overpic}
    \caption{(\textbf{Left}) : Angular resolution. (\textbf{Right}) : Energy resolution. On each plot, the single model is plotted with the multi models closest to that testing direction. The RF curves are taken from~\cite{Abe_2023}.}
    \label{fig:res}
\end{figure}

A quantitative comparison of model performance is presented in Figure \ref{fig:res}. The analysis reveals a clear performance trade-off between the two approaches. The Multi-Model strategy yields a notable improvement in energy resolution, particularly at energies below \textasciitilde500 GeV. This is a critical advantage for spectral analyses. In contrast, the angular resolution of the Single Model is better across the full energy range. Given the significant overhead of managing multiple models, the Single Model's ability to achieve similar angular resolution makes it a compelling choice for analyses of large amount of data spanning across zeniths.

A quantitative comparison of model performance is presented in Figure \ref{fig:res}. The analysis reveals that the multi-model approach yields a notable improvement in energy resolution, particularly at lower energies below \textasciitilde500 GeV, when compared to the single-model, more suitable for higher energies. While the angular resolution (68\% containment radius) is broadly comparable between the models, the single-model consistently performs better, due to a broader training sample. Note that at lower energies, the random forest implements a cut of the events where the shower development direction (disp sign in \cite{Abe_2023}) reconstruction fails, artificially improving the resolution curves at lower energies. CTLearn does not use disp sign reconstruction and can therefore not apply such a cut.

\section{Crab nebula data}
\begin{figure}
    \centering
    \begin{overpic}[width=15cm]{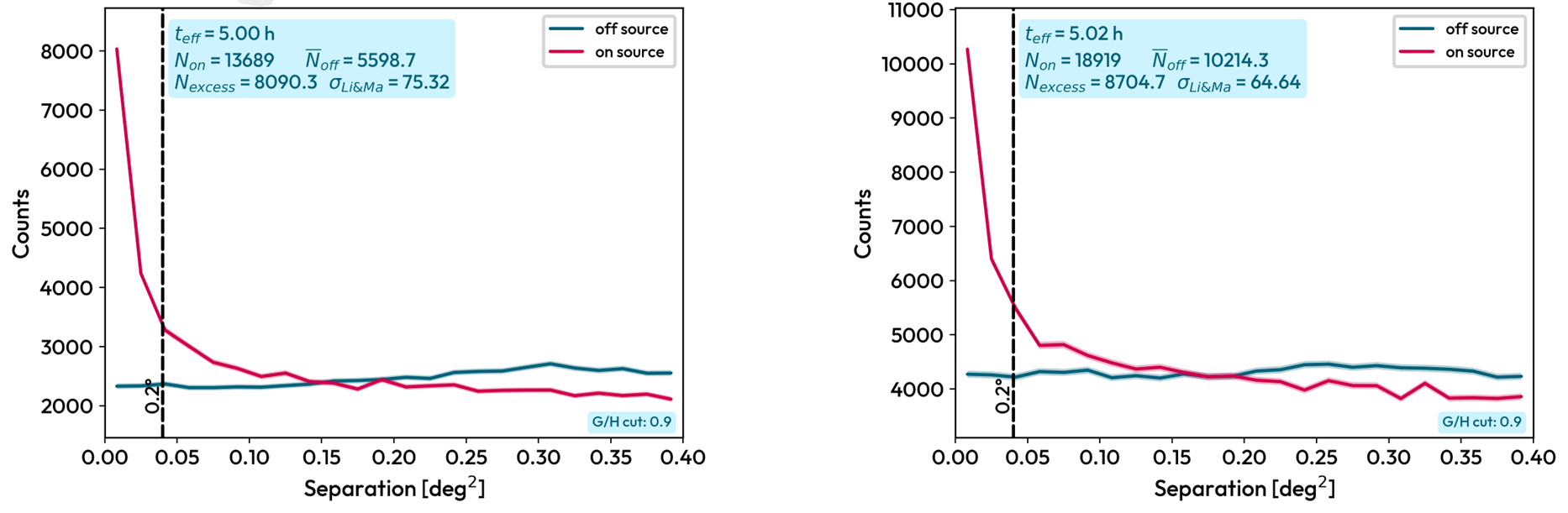}
        \put(75,20){\textcolor{gray}{Preliminary}}
        \put(20,20){\textcolor{gray}{Preliminary}}
    \end{overpic}
    \caption{Theta square plots of the LST-1 Crab nebula data. (\textbf{Left}) : Single model. (\textbf{Right}) : Multi models.}
    \label{fig:theta2}
\end{figure}

The models were subsequently applied to 5 hours of LST-1 observations of the Crab Nebula. The resulting theta-squared plots (Figure \ref{fig:theta2}) show a clear excess of events from the source direction for both the single and multi-model approaches, confirming a significant detection. 

For real data analysis, the same cuts that were optimized on MC are applied, to produce the desired metrics. Note that the LST-1 Crab performance paper~\cite{Abe_2023} does not use the same optimization algorithm on real data for benchmarking of RF models, therefore, a direct comparison of the metrics computed here with those in that work would not be appropriate. It is the reason the RF curves are not shown on some plots. Plans to compare to the RF method are described in the outlook.

\begin{figure}
    \centering
    \begin{overpic}[width=7cm]{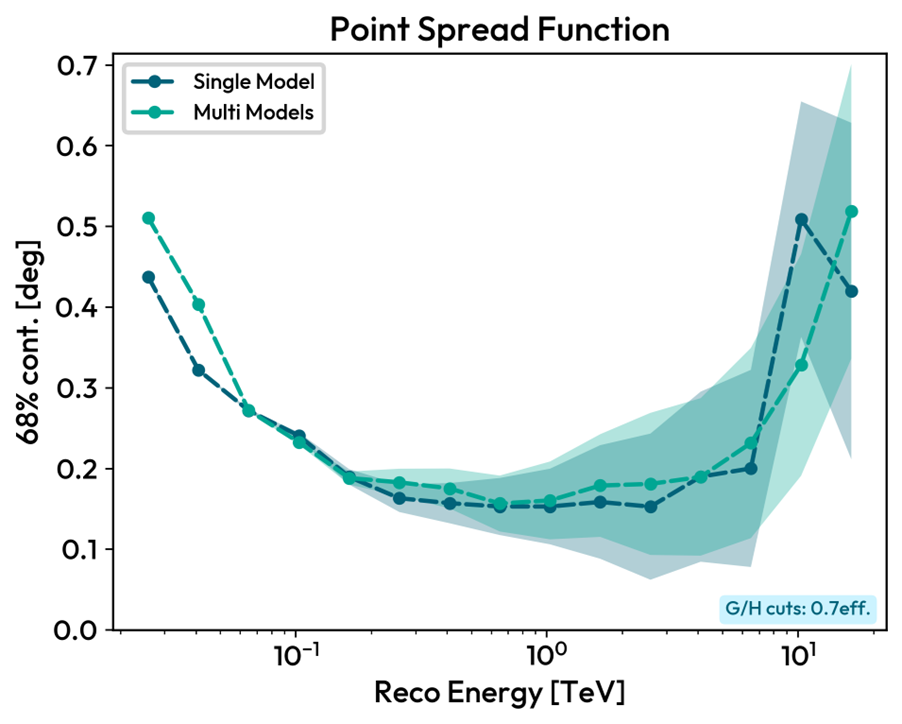}
        \put(45,50){\textcolor{gray}{Preliminary}}
    \end{overpic}
    \begin{overpic}[width=7.5cm]{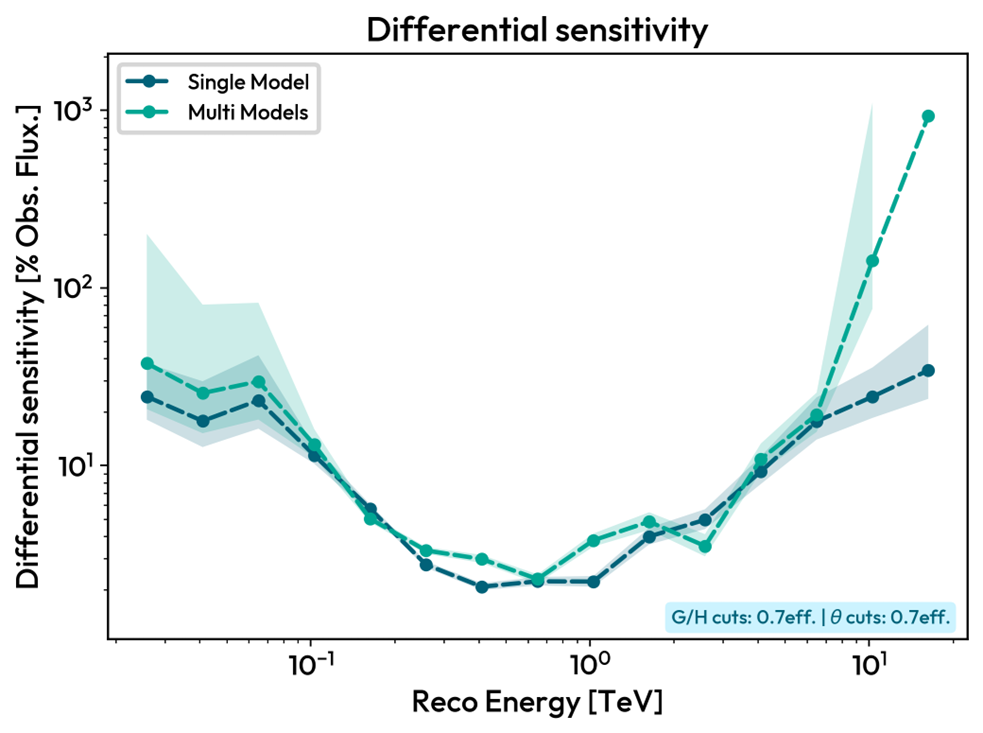}
        \put(45,45){\textcolor{gray}{Preliminary}}
    \end{overpic}
    \caption{(\textbf{Left}) : Point Spread Function computed on LST-1 Crab nebula data, comparing the single model to the multi model approach. (\textbf{Right}) : Differential sensitivity obtained with the analyzed sample in units of the total observed flux.}
    \label{fig:psf_sens}
\end{figure}

In order to evaluate what model strategy is the best for sensitivity and real data analysis, we compute the following metrics :

\begin{enumerate}

  \item Point Spread Function (PSF)

The PSF on real data is simply computes as the 68\% containment radius of the events around the source, in bins of reconstructed energy. For this analysis, gammaness cuts for 70\% are applied, but not the theta cuts (Figure \ref{fig:psf_sens}).
  \item Sensitivity 

On real data, sensitivity is computed by first applying the MC-optimized cuts for 70\% efficiency on the data. Then, the goal is to find the minimum flux fraction that results in a $5\sigma$ detection in 50h, compared to the actual sensitivity and observation time of the analyzed sample.
For this, on proceeds iteratively, setting the current observed flux factor to $f_{f}^0=1$ and scaling the flux by the ratio of the target observation time (50h) and effective observation time $f_t=50h/t_{eff}$. Then, computing the corresponding Li\&Ma significance, and scaling the flux by the ratio of the target significance to the current one ($5\sigma$). Until the significance of the scaled sample gives 5 sigmas, the flux factor in percentage of observed flux is the sensitivity on real data (Figure \ref{fig:psf_sens}).



\end{enumerate}

Crucially, the high-level performance metrics computed from this data (Figure \ref{fig:res}) validate the trends observed in simulations. The PSF, or 68\% containment radius, is comparable between the two models. However, the differential sensitivity measurement shows a clear advantage for the single-model approach, which achieves a better sensitivity across the majority of the energy range. This indicates that the single-model strategy would allow for the detection of fainter sources in an equivalent amount of observation time.

\section{Discussion and conclusion}

In this work, we evaluated two deep-learning strategies for LST-1 event reconstruction using CTLearn: a single, universal model and a set of ten specialized multi-models. Our validation on both Monte-Carlo simulations and 5 hours of Crab Nebula observations demonstrated that while the specialized Multi-Model approach offers superior energy resolution, the operationally simpler Single Model achieves highly comparable angular resolution and, most critically, better overall differential sensitivity. Since sensitivity is the key metric for the detection of faint gamma-ray sources, the single-model approach currently stands as the preferred strategy for general-purpose analysis.

These results suggest a promising path for future development: a hybrid reconstruction method. Such a system could leverage the strengths of both approaches by using the multi-model output exclusively for energy estimation, while employing the more sensitive single model for event direction and gamma/hadron classification. This could provide optimal performance for all reconstruction tasks simultaneously.

A limitation of this study is the lack of a direct benchmark against the standard Random Forest (RF)~\cite{2008NIMPA.588..424A} analysis pipeline. Future work will focus on this essential comparison. Additionally, we plan to explore the use of transfer learning to drastically reduce the training time of specialized models, which would make the development of a hybrid system more efficient. In conclusion, CTLearn provides a powerful and flexible framework for CTAO data analysis, and these results provide a foundation for developing advanced, highly-optimized reconstruction techniques.

\bibliographystyle{unsrturl}
\bibliography{biblio.bib}

\textbf{Full Author List: CTAO-LST Project}

\tiny{\noindent
K.~Abe$^{1}$,
S.~Abe$^{2}$,
A.~Abhishek$^{3}$,
F.~Acero$^{4,5}$,
A.~Aguasca-Cabot$^{6}$,
I.~Agudo$^{7}$,
C.~Alispach$^{8}$,
D.~Ambrosino$^{9}$,
F.~Ambrosino$^{10}$,
L.~A.~Antonelli$^{10}$
% ...
The CTAO-LST Project
}\\

\tiny{\noindent$^{1}${Department of Physics, Tokai University, 4-1-1, Kita-Kaname, Hiratsuka, Kanagawa 259-1292, Japan}.
$^{2}${Institute for Cosmic Ray Research, University of Tokyo, 5-1-5, Kashiwa-no-ha, Kashiwa, Chiba 277-8582, Japan}.
$^{3}${INFN and Università degli Studi di Siena, Dipartimento di Scienze Fisiche, della Terra e dell'Ambiente (DSFTA), Sezione di Fisica, Via Roma 56, 53100 Siena, Italy}.
$^{4}${Université Paris-Saclay, Université Paris Cité, CEA, CNRS, AIM, F-91191 Gif-sur-Yvette Cedex, France}.
$^{5}${FSLAC IRL 2009, CNRS/IAC, La Laguna, Tenerife, Spain}.
$^{6}${Departament de Física Quàntica i Astrofísica, Institut de Ciències del Cosmos, Universitat de Barcelona, IEEC-UB, Martí i Franquès, 1, 08028, Barcelona, Spain}.
$^{7}${Instituto de Astrofísica de Andalucía-CSIC, Glorieta de la Astronomía s/n, 18008, Granada, Spain}.
$^{8}${Department of Astronomy, University of Geneva, Chemin d'Ecogia 16, CH-1290 Versoix, Switzerland}.
$^{9}${INFN Sezione di Napoli, Via Cintia, ed. G, 80126 Napoli, Italy}.
$^{10}${INAF - Osservatorio Astronomico di Roma, Via di Frascati 33, 00040, Monteporzio Catone, Italy}.
% ...
}

\acknowledgments 
\tiny{
We gratefully acknowledge financial support from the following agencies and organisations:
Conselho Nacional de Desenvolvimento Cient\'{\i}fico e Tecnol\'{o}gico (CNPq), Funda\c{c}\~{a}o de Amparo \`{a} Pesquisa do Estado do Rio de Janeiro (FAPERJ), Funda\c{c}\~{a}o de Amparo \`{a} Pesquisa do Estado de S\~{a}o Paulo (FAPESP), Funda\c{c}\~{a}o de Apoio \`{a} Ci\^encia, Tecnologia e Inova\c{c}\~{a}o do Paran\'a - Funda\c{c}\~{a}o Arauc\'aria, Ministry of Science, Technology, Innovations and Communications (MCTIC), Brasil;
Ministry of Education and Science, National RI Roadmap Project DO1-153/28.08.2018, Bulgaria;
Croatian Science Foundation (HrZZ) Project IP-2022-10-4595, Rudjer Boskovic Institute, University of Osijek, University of Rijeka, University of Split, Faculty of Electrical Engineering, Mechanical Engineering and Naval Architecture, University of Zagreb, Faculty of Electrical Engineering and Computing, Croatia;
Ministry of Education, Youth and Sports, MEYS  LM2023047, EU/MEYS CZ.02.1.01/0.0/0.0/16\_013/0001403, CZ.02.1.01/0.0/0.0/18\_046/0016007, CZ.02.1.01/0.0/0.0/16\_019/0000754, CZ.02.01.01/00/22\_008/0004632 and CZ.02.01.01/00/23\_015/0008197 Czech Republic;
CNRS-IN2P3, the French Programme d’investissements d’avenir and the Enigmass Labex, 
This work has been done thanks to the facilities offered by the Univ. Savoie Mont Blanc - CNRS/IN2P3 MUST computing center, France;
Max Planck Society, German Bundesministerium f{\"u}r Bildung und Forschung (Verbundforschung / ErUM), Deutsche Forschungsgemeinschaft (SFBs 876 and 1491), Germany;
Istituto Nazionale di Astrofisica (INAF), Istituto Nazionale di Fisica Nucleare (INFN), Italian Ministry for University and Research (MUR), and the financial support from the European Union -- Next Generation EU under the project IR0000012 - CTA+ (CUP C53C22000430006), announcement N.3264 on 28/12/2021: ``Rafforzamento e creazione di IR nell’ambito del Piano Nazionale di Ripresa e Resilienza (PNRR)'';
ICRR, University of Tokyo, JSPS, MEXT, Japan;
JST SPRING - JPMJSP2108;
Narodowe Centrum Nauki, grant number 2023/50/A/ST9/00254, Poland;
The Spanish groups acknowledge the Spanish Ministry of Science and Innovation and the Spanish Research State Agency (AEI) through the government budget lines
PGE2022/28.06.000X.711.04,
28.06.000X.411.01 and 28.06.000X.711.04 of PGE 2023, 2024 and 2025,
and grants PID2019-104114RB-C31,  PID2019-107847RB-C44, PID2019-104114RB-C32, PID2019-105510GB-C31, PID2019-104114RB-C33, PID2019-107847RB-C43, PID2019-107847RB-C42, PID2019-107988GB-C22, PID2021-124581OB-I00, PID2021-125331NB-I00, PID2022-136828NB-C41, PID2022-137810NB-C22, PID2022-138172NB-C41, PID2022-138172NB-C42, PID2022-138172NB-C43, PID2022-139117NB-C41, PID2022-139117NB-C42, PID2022-139117NB-C43, PID2022-139117NB-C44, PID2022-136828NB-C42, PDC2023-145839-I00 funded by the Spanish MCIN/AEI/10.13039/501100011033 and “and by ERDF/EU and NextGenerationEU PRTR; the "Centro de Excelencia Severo Ochoa" program through grants no. CEX2019-000920-S, CEX2020-001007-S, CEX2021-001131-S; the "Unidad de Excelencia Mar\'ia de Maeztu" program through grants no. CEX2019-000918-M, CEX2020-001058-M; the "Ram\'on y Cajal" program through grants RYC2021-032991-I  funded by MICIN/AEI/10.13039/501100011033 and the European Union “NextGenerationEU”/PRTR and RYC2020-028639-I; the "Juan de la Cierva-Incorporaci\'on" program through grant no. IJC2019-040315-I and "Juan de la Cierva-formaci\'on"' through grant JDC2022-049705-I. They also acknowledge the "Atracci\'on de Talento" program of Comunidad de Madrid through grant no. 2019-T2/TIC-12900; the project "Tecnolog\'ias avanzadas para la exploraci\'on del universo y sus componentes" (PR47/21 TAU), funded by Comunidad de Madrid, by the Recovery, Transformation and Resilience Plan from the Spanish State, and by NextGenerationEU from the European Union through the Recovery and Resilience Facility; “MAD4SPACE: Desarrollo de tecnolog\'ias habilitadoras para estudios del espacio en la Comunidad de Madrid" (TEC-2024/TEC-182) project funded by Comunidad de Madrid; the La Caixa Banking Foundation, grant no. LCF/BQ/PI21/11830030; Junta de Andaluc\'ia under Plan Complementario de I+D+I (Ref. AST22\_0001) and Plan Andaluz de Investigaci\'on, Desarrollo e Innovaci\'on as research group FQM-322; Project ref. AST22\_00001\_9 with funding from NextGenerationEU funds; the “Ministerio de Ciencia, Innovaci\'on y Universidades”  and its “Plan de Recuperaci\'on, Transformaci\'on y Resiliencia”; “Consejer\'ia de Universidad, Investigaci\'on e Innovaci\'on” of the regional government of Andaluc\'ia and “Consejo Superior de Investigaciones Cient\'ificas”, Grant CNS2023-144504 funded by MICIU/AEI/10.13039/501100011033 and by the European Union NextGenerationEU/PRTR,  the European Union's Recovery and Resilience Facility-Next Generation, in the framework of the General Invitation of the Spanish Government’s public business entity Red.es to participate in talent attraction and retention programmes within Investment 4 of Component 19 of the Recovery, Transformation and Resilience Plan; Junta de Andaluc\'{\i}a under Plan Complementario de I+D+I (Ref. AST22\_00001), Plan Andaluz de Investigaci\'on, Desarrollo e Innovación (Ref. FQM-322). ``Programa Operativo de Crecimiento Inteligente" FEDER 2014-2020 (Ref.~ESFRI-2017-IAC-12), Ministerio de Ciencia e Innovaci\'on, 15\% co-financed by Consejer\'ia de Econom\'ia, Industria, Comercio y Conocimiento del Gobierno de Canarias; the "CERCA" program and the grants 2021SGR00426 and 2021SGR00679, all funded by the Generalitat de Catalunya; and the European Union's NextGenerationEU (PRTR-C17.I1). This research used the computing and storage resources provided by the Port d’Informaci\'o Cient\'ifica (PIC) data center.
State Secretariat for Education, Research and Innovation (SERI) and Swiss National Science Foundation (SNSF), Switzerland;
The research leading to these results has received funding from the European Union's Seventh Framework Programme (FP7/2007-2013) under grant agreements No~262053 and No~317446;
This project is receiving funding from the European Union's Horizon 2020 research and innovation programs under agreement No~676134;
ESCAPE - The European Science Cluster of Astronomy \& Particle Physics ESFRI Research Infrastructures has received funding from the European Union’s Horizon 2020 research and innovation programme under Grant Agreement no. 824064.}

\end{document}